\documentclass[pra,twocolumn]{revtex4-1}
\usepackage{amsmath}
\usepackage{amssymb}
\usepackage{graphicx}
\usepackage{color}

\def\braket#1{\left\langle#1\right\rangle}
\def\I{\mathcal{I}}
\def\N{\mathcal{N}}

\def\IN{\textrm{in}}

\begin{document}

\title{Noise Properties of Coherent Perfect Absorbers and Critically-coupled Resonators}

\author{Y.~D.~Chong}
\email{yidong@ntu.edu.sg}

\affiliation{Division of Physics and Applied Physics, School of Physical and Mathematical Sciences, Nanyang Technological University, Singapore 637371, Singapore}

\author{Hui Cao}

\author{A.~D.~Stone}

\affiliation{Department of Applied Physics, Yale University, New
  Haven, Connecticut 06520}

\begin{abstract}
The performance of a coherent perfect absorber (time-reversed laser)
is limited by quantum and thermal noise.  At zero temperature, the
quantum shot noise dominates the signal for frequencies close to the
resonance frequency, and both vanish exactly at the resonance
frequency.  We compute the sensitivity of the absorbing cavity as a
background-free detector, limited by finite signal or detector
bandwidth.
\end{abstract}


\maketitle

In recent work the authors and collaborators have proposed
\cite{chong10} and demonstrated \cite{wan11} the phenomenon of
coherent perfect absorption, or ``time-reversed lasing''. Applying the
time-reversal operation to the classical electromagnetic equations
yields the following statement: If a cavity containing a gain medium
characterized by an amplifying refractive index $n(\vec{r}) = n_1
(\vec{r}) - i n_2 (\vec{r})$ reaches the lasing threshold for a
certain lasing mode at frequency $ \omega_0$, then a cavity containing
a dissipative refractive index $n^*(\vec{r})$ will perfectly absorb an
input mode at $\omega_0$ corresponding to the time-reverse of the
lasing mode.  We refer to such a lossy cavity as a coherent perfect
absorber (CPA).  Assuming the input signal is perfectly monochromatic,
the above ``CPA theorem'' is rigorously true within classical
electromagnetic theory, where the effects of quantum and thermal noise
are neglected.  The CPA is a generalization to arbitrary geometries
and arbitrary numbers of input channels of the well-known concept of a
critically-coupled resonator (CCR) \cite{ccr}, an optical device which
can be used for switching, modulation, enhanced photodetection and
sensing \cite{afp,rce}, and which may be regarded as the single
channel limit of a CPA.  Because a CPA is associated with a vanishing
output signal in the classical zero-temperature limit, a CPA (or CCR)
can function as a background-free detector or interferometer, similar
to a Mach-Zender interferometer (MZ).  The fundamental limits to its
effectiveness in this role are determined by quantum and thermal
noise, which are the subject of the present paper.


Quantum fluctuations, such as spontaneous emission, break the symmetry
between emission and absorption.  Hence, noise processes in a CPA (or
CCR) differ from those in a laser or amplifier.  Within semiclassical
theory, the laser has zero linewidth; by including quantum
fluctuations, one obtains the Schawlow-Townes (ST) linewidth
\cite{ST}, which decreases inversely with the output power well above
threshold.  As is well-known, the ST linewidth arises from the
dephasing of the above-threshold laser field due to quantum noise
(usually characterized as ``one noise photon per relaxation time per
mode'').  Because a CPA does not contain an inverted medium like a
laser, and spontaneous emission vanishes at low temperature, there is
no direct analog of the ST linewidth in a CPA.  At zero temperature,
the frequency characteristics of the output are determined entirely by
the input, and the CPA absorption resonance has approximately twice
the passive cavity linewidth, as expected for critical coupling.
However, this linewidth is further modified, as we shall show below,
by an analog of the Petermann factor \cite{Petermann,Haus}; this
effect has not, to our knowledge, been recognized in absorbing
systems.

Due to the lack of spontaneous emission at zero temperature, were it
possible to probe the resonance with a truly monochromatic input at
the CPA resonance, the perfect absorption predicted by the classical
theory would occur---even with quantum fluctutations taken into
account.  In this situation, the quantum shot noise vanishes
simultaneously with the zero of the average output.  In practice,
however, the unavoidable linewidth of the input field, $\Delta_{\IN}$,
combines with quantum shot noise to generate a finite noise floor even
at the resonance frequency.  The noise dominates the signal within an
interval $\delta\omega_{\textrm{x}} \propto
\sqrt{\Delta_{\IN}/P_{\IN}}$ around the resonance frequency, where
$P_{\IN}$ is the power of the input signal.  This behavior is similar
to a MZ, with the absorption into an external reservoir playing the
role of an unobserved MZ output port.  However, in contrast to the MZ,
the resonant frequency response of the cavity can give rise to
parametrically better signal-to-noise ratio in the vicinity of the
background-free point \cite{Chow}.  For finite temperature, the output
also includes thermal noise; at the absorption resonance, the thermal
emissivity takes the black-body value for a one-port system (CCR), but
can be significantly reduced in a multi-port CPA.

We can analyze the noise properties of the CPA using the standard
input-output framework of quantum optics \cite{Caves,beenakker0}.
Input and output photon operators are connected by the classical
electromagnetic scattering matrix $S_{ij} (\omega)$, where $j =
1,2,\dots,N$ denote the scattering channels.  The $S$-matrix for a CPA is sub-unitary
due to the presence of an absorbing reservoir, and describes a ``gray-body", with some
fraction of the incident photons absorbed, and the remainder scattered out at the input 
frequency. The scattering channels denote different spatial states in the asymptotic region 
which suffice to represent an arbitrary incoming or outgoing field at $\omega$, e.g. incoming and outgoing
angular momentum channels in a two-dimensional scattering geometry.
The input photon operators are denoted by $a_j$, and the output photon operators by
$b_j$.  The special case of a single scattering channel ($N=1$)
corresponds to a standard CCR.  We assume that the CPA is coupled to
an ideal external reservoir, so that the absorption of light produces
negligible heating and hence negligible change in the output noise.
The photon operators are related by an input/output relation
\cite{beenakker0,Caves,Loudon}
\begin{equation}
  b_i(\omega) = \sum_j S_{ij}(\omega) \, a_j(\omega) + \sum_\nu
  U_{i\nu}(\omega)\, c_\nu(\omega),
  \label{input output}
\end{equation}
where the $c_{\nu}$'s are ladder operators for reservoir quanta.  Here
we have assumed a coupling to the reservoir which adds the minimum
amount of quantum noise \cite{Clerk}.  The requirement that $a$, $b$,
and $c$ obey canonical commutation relations,
e.g.~$[a_i(\omega),a_j^\dagger(\omega')] =
\delta_{ij}\,\delta(\omega-\omega')$, yields the
fluctuation-dissipation relation \cite{beenakker0}
\begin{equation}
  SS^\dagger + UU^\dagger = \mathbf{1},
  \label{fluct-diss}
\end{equation}
where $\mathbf{1}$ is the $N\times N$ identity matrix.  Equation
(\ref{fluct-diss}) generalizes the unitarity relation of the lossless
system, and implies that the eigenvalues of the $S$-matrix generically
have magnitude smaller than unity.

We are interested in the shot noise in the output field, for a
coherent input at some frequency $\omega$.  To work with
equal-frequency correlators, it is convenient to rescale the continuum
operators $\{a_i,b_i,c_\nu\}$ to discrete operators
$\{\hat{a}_i,\hat{b}_i,\hat{c}_\nu\}$, which are normalized so that
the equal-frequency commutator is unity,
e.g.~$[\hat{a}_i(\omega),\hat{a}_j^\dagger(\omega)] = \delta_{ij}$.  Here and
in the following, we omit the $\omega$ dependence from the notation.
Next, the input photon operator can be re-written using a displacement
transformation \cite{Lagendijk,Clerk}
\begin{equation}
  \hat{a}_i = \alpha_i + \hat{a}_i',
\end{equation}
where $\alpha_i$ is a coherent state amplitude, and $\hat{a}_i'$ is an
operator accounting for fluctuations around the coherent state, which
likewise obeys the canonical commutation relation
$[\hat{a}_i',\hat{a}_j'^\dagger] = \delta_{ij}$.  Hence,
\begin{equation}
  \hat{b}_i = \sum_j S_{ij}\,(\hat{\alpha}_j + \hat{a}_j') + \sum_\nu
  U_{i\nu}\,\hat{c}_\nu.
  \label{bout}
\end{equation}

The operator $\N_i \equiv \hat{b}_i^\dagger\hat{b}_i$ gives the output
photon flux per unit frequency (at frequency $\omega$) in channel $i$.
(Correspondingly, $\I_i = \hbar\omega_0 \N_i$ describes the spectral
density.)  Using (\ref{bout}), we can calculate the expectation value
and correlation function for $\N_i$, using standard Gaussian $n$-point
operator correlators \cite{Lagendijk}.  We take $\langle
\hat{a}_i'\rangle = \langle \hat{c}_\nu \rangle = 0$, and $\langle
\hat{a}'^{\dagger}_i \hat{a}_j'\rangle = 0$ (zero net fluctuation
around the specified coherent input amplitude), and $\langle
\hat{c}^\dagger_\mu \hat{c}_\nu\rangle = \delta_{\mu\nu} \,f(T)$ where
$f(T) = [\exp(\hbar\omega/k_BT)-1]^{-1}$ and $T$ is the temperature of
the reservoir.  The result for $\braket{\N_i}$ is
\begin{equation}
  \braket{\N_i} = |(S\mathbf{\alpha})_i|^2 +
  \left[1-(SS^\dagger)_{ii}\right] \,f(T).
  \label{a1}
\end{equation}
Thus the total output is
\begin{equation}
  \braket{\N} = \sum_i \braket{\N_i}= |S\alpha|^2 + f(T)\,
  \textrm{Tr}(\mathbf{1} - SS^\dagger),
  \label{total N output}
\end{equation}
and the noise is
\begin{align}
  \begin{aligned}
    \langle\delta & \N^2 \rangle \equiv \sum_{ij}
    \left[\braket{\N_i\,\N_j} - \braket{\N_i}\braket{\N_j}\right]\\ &=
    |S\alpha|^2 \\ &\quad + f(T) \, \Big\{2|S\alpha|^2 - 2
    |S^\dagger S \alpha|^2 + \textrm{Tr}\left(\mathbf{1} -
    SS^\dagger\right) \Big\} \\ &\quad + [f(T)]^2 \;
    \textrm{Tr}\left[\left(\mathbf{1}-SS^\dagger\right)^2\right].
    \label{output noise}
  \end{aligned}
\end{align}
For $T \to 0, f(T) \to 0$, (\ref{total N output}) and (\ref{output noise}) reduce to
the Poissonian result
\begin{equation}
  \braket{\N} = \braket{\delta\N^2} = |S\alpha|^2.
  \label{poisson}
\end{equation}
The CPA condition is achieved when the index of refraction of the
cavity, $n(\vec{r})$, is chosen such that there exists an eigenvector
of the $S$-matrix with eigenvalue zero at a specific input frequency
$\omega_0$.  If $\alpha$ is chosen to be this eigenvector, then we see
from Eq.~(\ref{poisson}) that the mean outgoing photon flux vanishes
(as it should be from the classical CPA theorem), and so does its
variance (the shot noise).  This confirms the statement that the CPA
effect is unaffected by quantum fluctuations at zero temperature for a
purely monochromatic input.

At finite temperature, the shot noise dominated regime corresponds to $T \ll \hbar \omega_0 / k_B$,  For $T \neq 0$, $\braket{\N}$ and
$\braket{\delta\N^2}$ are unequal, and differ by a term arising from the beating
between the thermal emission and the scattered flux:
\begin{equation}
  \braket{\delta\N^2} - \braket{\N} = f(T) \left[2|S\alpha|^2 - 2
    |S^\dagger S \alpha|^2\right] + O(f^2).
  \label{variance vs mean}
\end{equation}
This difference can be shown to be strictly positive for any input field. The term linear in $f(T)$ vanishes 
when the input field, $\alpha$, corresponds to the zero eigenvector, so thermal noise in the CPA is very small in the 
low temperature regime.

\begin{figure}
\centering
\includegraphics[width=0.47\textwidth]{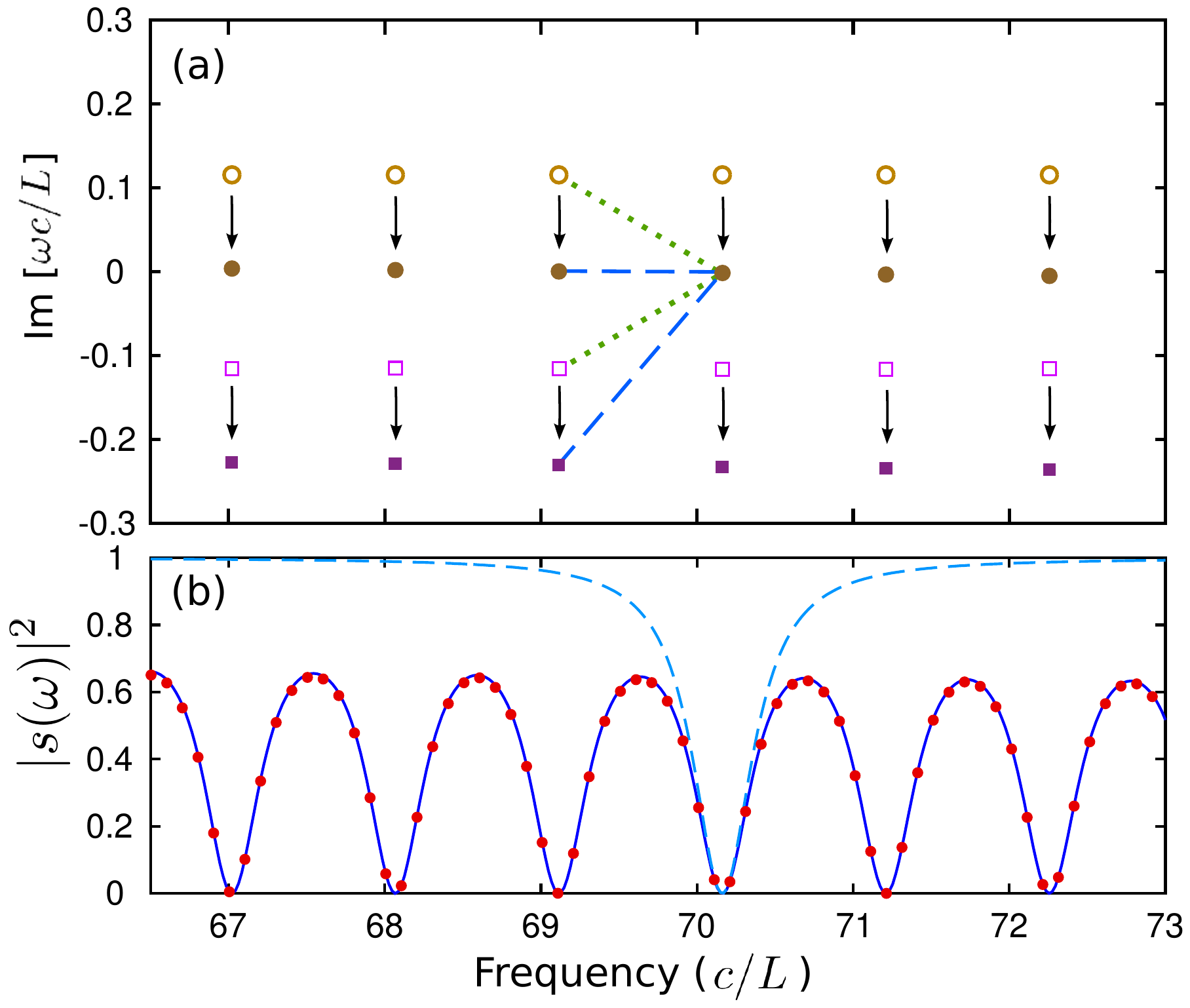}
\caption{(color online) Scattering properties of a one-port coherent
  perfect absorber (critically-coupled resonator) consisting of a
  one-dimensional uniform dielectric slab of length $L$, with a
  perfect mirror on one side and vacuum on the other.  (a) Location of
  poles (squares) and zeros (circles) in the complex frequency plane.
  Open symbols show the poles and zeros for the passive cavity with
  refractive index $n = 3$, and filled symbols for an absorbing cavity
  with refractive index $n = 3 + 0.005i$.  The dashed and dotted lines
  are guides to the eye for the geometric interpretation of the
  Petermann factor given in Eq.~(\ref{s omega 2}).  Without
  absorption, $|\omega - \omega^z| = |\omega - \omega^p|$, where
  $\omega \in \mathbb{R}$ and $\omega^z$ and $\omega^p$ are the
  frequencies of a neighboring pair of zeros and poles (green dots).
  With very narrow band absorption, near a CPA resonance (in this case
  $\omega \approx 70.1 c/L$), this remains approximately true.
  However, for broadband absorption, $|\omega - \omega^z| < |\omega -
  \omega^p|$ (blue dashes).  (b) Plot of $|s(\omega)|^2$, where
  $s(\omega)$ is the eigenvalue of the scattering matrix (i.e.~in this
  one-port case the reflection coefficient).  The red circles show
  exact numeric results obtained by the transfer matrix method, while
  the solid blue curve shows Eq.~(\ref{s omega 2}), with the product
  taken over the 20 pairs of poles and zeros nearest to $\omega = 70
  L/c$.  The dashed curve shows the single-resonance approximation,
  Eq.~(\ref{s omega}), using the pole and zero nearest to $\omega = 70
  L/c$. }
\label{fig_poleszeros}
\end{figure}

To estimate the sensitivity of a CPA as a detector, we study its
behavior for frequencies near the perfect absorption resonance,
$\omega_0$.  For now, we consider $T = 0$.  In the absence of loss,
the $S$-matrix has a sequence of resonances, associated with poles in
the lower half of the complex $\omega$ plane and symmetrically placed
zeros in the upper half plane \cite{chong10}, at discrete frequencies
$\omega = \omega_0 \mp i \gamma_c/2$.  Near $\omega_0$, one of the
$S$-matrix eigenvalues takes the following approximate form
\cite{Walls}:
\begin{equation}
  s(\omega) \approx e^{i\varphi(\omega)} \; \frac{\omega - \omega_0 -
    i\gamma_c/2}{\omega - \omega_0 + i\gamma_c/2},
  \label{sapprox}
\end{equation}
where $\varphi$ is an irrelevant phase factor and $\gamma_c$ is the
cavity linewidth (FWHM).  Eq.~(\ref{sapprox}) is a ``single resonance
approximation'' which ignores the presence of other nearby poles and
zeros.  It satisfies the requirements that $|s| = 1$ when $\omega \in
\mathbb{R}$ (the lossless $S$-matrix has unimodular eigenvalues), and
that $s$ goes to zero and infinity at $\omega = \omega_0 \pm i
\gamma_c/2$.  Adding absorption pushes the zero and pole down in the
complex frequency plane, as shown in Fig.~\ref{fig_poleszeros}(a); to
lowest order, they move down by equal amounts.  To achieve the CPA
condition, exactly enough absorption is added to push the zero down to
the real $\omega$ axis.  Then the eigenvalue of the absorbing cavity
is
\begin{equation}
  s(\omega) \approx e^{i\varphi(\omega)} \; \frac{\omega - \omega_0 +
    i \,\delta\gamma}{\omega - \omega_0 + i\gamma_c},
  \label{s omega}
\end{equation}
where the parameter $\delta\gamma \ll \gamma_c$ represents a small
detuning of the material loss from the perfect absorption resonance.
Note that the resonance of the absorbing cavity has twice the width of
the passive cavity resonance ($\gamma_c/2 \to \gamma_c$).  This is due
to the ``critical coupling'' condition that the absorption loss rate
is equal to the scattering loss rate.

In this case, if the input mode, $\alpha$, corresponds to the zero
eigenvector of the $S$-matrix, then (at $T=0$):
\begin{equation}
  \braket{\N} = \braket{\delta\N^2} \approx \frac{(\omega-\omega_0)^2
    + \delta\gamma^2}{(\omega - \omega_0)^2 + \gamma_c^2} \;
  \frac{P_{\IN}(\omega)}{\hbar \omega \Delta_{\IN}}.
  \label{T=0 CPA}
\end{equation}
Here we have expressed the input photon flux per
unit frequency, $|\alpha|^2$, in terms of the input power
\begin{equation}
  P_{\IN}(\omega) = |\alpha(\omega)|^2 \hbar \omega \Delta_{\IN}.
\end{equation}
This power is averaged over some bandwidth $\Delta_{\IN}$, which is
taken to be the frequency resolution of either the input state or the
output detector, whichever is smaller.

\begin{figure}
\centering
\includegraphics[width=0.4\textwidth]{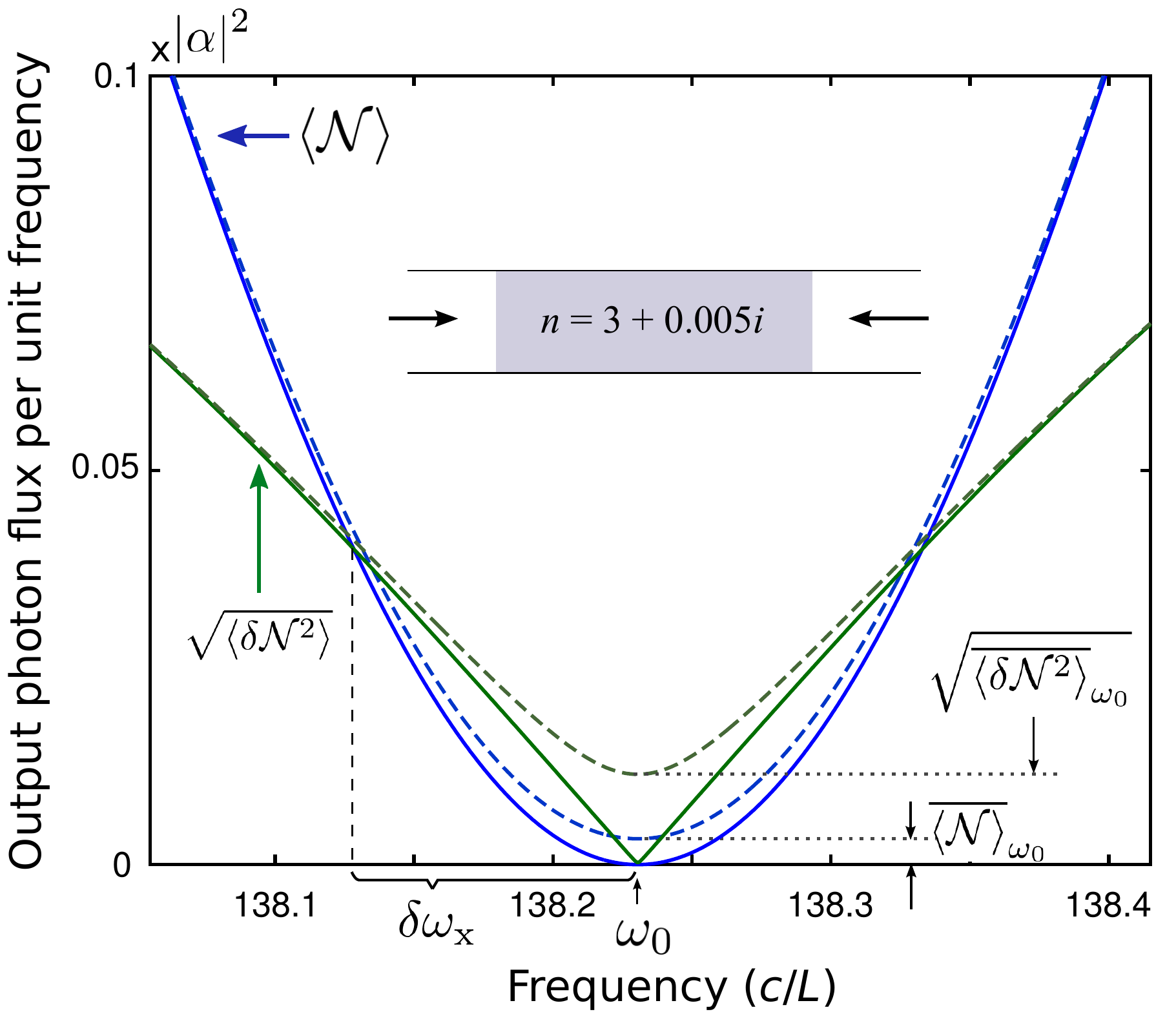}
\caption{(color online) Plot mean output power ($\mathcal{N}$) and
  shot noise power ($\sqrt{\delta\N^2}$), as a function of frequency
  near a CPA zero, for $T=0$.  The cavity is a two-sided uniform
  dielectric slab with refractive index $3 + 0.005i$ and length $L$
  (see schematic).  For this case the CPA eigenmodes are simply
  incoming coherent plane waves of equal amplitude from each
  direction, with either even or odd parity.  Results for an even CPA
  mode at frequency $\omega \approx 138.23 \, c/L$ are shown.  We have
  chosen the input intensity $|\alpha|^2 = 25$.  The solid curves are
  obtained from Eq.~(\ref{poisson}), using the transfer matrix method
  to find $S(\omega)$.  The dashed curves are obtained by averaging
  these values over a bandwidth $\Delta_{\textrm{in}} = 0.1 \, c/L$,
  corresponding to a finite spectrometer resolution.  Inset: schematic
  of the system.}
\label{fig_zerot}
\end{figure}

According to Eq.~(\ref{T=0 CPA}), as $\delta \gamma \to 0$ and $\omega
\to \omega_0$, the mean and variance of the photon flux behaves as
\begin{equation}
  \braket{\N} = \braket{\delta\N^2} \approx \frac{(\omega-\omega_0)^2
}{ \gamma_c^2} \;
  \frac{P_{\IN}(\omega)}{\hbar \omega \Delta_{\IN}}.
  \label{spa}
\end{equation}
The CPA resonance has a quadratic zero at $\omega_0$
(Fig.~\ref{fig_zerot}), and, in the single resonance approximation,
its width is $ \gamma_c$, the critically-coupled cavity linewidth. (We
assume that $\gamma_c $ is much less than the free spectral range of
the resonator).

However, adding a lossy medium to the resonator affects the $S$-matrix
eigenvalue beyond the single-resonance approximation, and the
corrections to this approximation generically increase the width.
This is the exact analog of the Petermann factor in lasers, which
increases the ST linewidth \cite{Petermann,Haus}.  This is most easily
seen in the one-port case, where we can express the $S$-matrix
eigenvalue with a zero on the real axis as
\begin{equation}
  s(\omega) = e^{i\varphi(\omega)} \; \frac{\omega - \omega_0}{\omega
    - \omega_0 + i\gamma_c} \; \prod_n \frac{\omega -
    \omega^z_n}{\omega - \omega^p_n},
  \label{s omega 2}
\end{equation}
where $n$ indexes the other pairs of zeros and poles, which have
frequencies $\omega_n^z$ and $\omega_n^p$ respectively.  This ansatz
agrees very well with exact numerical calculations of $s(\omega)$, as
shown in Fig.~\ref{fig_poleszeros}(b).  Due to time-reversal symmetry,
in the absence of loss, each factor in the product would be
unimodular, since the zeros and poles are symmetrically distributed
around the real axis.  The introduction of loss breaks this symmetry
for all the poles and zeros, not just the zero at $\omega_0$; this
causes the zeros to move towards the real axis and the poles away, so
that $|\omega - \omega^z_n| < |\omega - \omega^p_n|$ for any real
$\omega$, as shown in Fig.~\ref{fig_poleszeros}(a). This effect is
neglected in the single resonance approximation of Eq.~(\ref{spa}).
Including it leads to
\begin{equation}
  \braket{\N} = \braket{\delta\N^2} \approx \frac{(\omega-\omega_0)^2
}{K \gamma_c^2} \;
  \frac{P_{\IN}(\omega)}{\hbar \omega \Delta_{\IN}},
  \label{sra}
\end{equation}
where $K = \prod_n |(\omega - \omega^p_n)/(\omega - \omega^z_n)| > 1$
for the one-channel case.  Calculating the Petermann factor for the
time-reversed counterpart of this cavity, which contains gain instead
of loss and sits at the first lasing threshold, would yield exactly
the same value of $K$ \cite{linewidth}.  This equivalence follows from
the properties of the $S$-matrix under time-reversal, as does the CPA
effect itself.  For a more complex, multichannel cavity, the explicit
calculation of the $S$-matrix eigenvalues is more complicated, but the
symmetry that leads to the factor $1/K$ in the eigenvalue still holds.
In other words, adding loss to the cavity so as to reach the CPA
condition increases the critically-coupled cavity resonance linewidth
from $\gamma_c \to \sqrt{K} \gamma_c$.

In calculating this broadening effect, one could assume that the loss
of the cavity is broadband, affecting all relevant poles and zeros
equally; a similar assumption of broadband gain is made in standard
calculations of the Petermann factor.  When this is not the case, both
for CPAs and lasers the Petermann correction is reduced, and needs to
be calculated using the frequency-dependent refractive index of the
cavity \cite{linewidth}.  For the subsequent analysis, we will assume
that the $K$ occurring in Eq.~(\ref{sra}) is a given parameter.

Exactly at $\omega_0$, both signal and noise vanish; near $\omega_0$,
the shot noise $[\braket{\delta\N^2}]^{1/2}$ dominates over the output
signal $\braket{\N}$.  The crossover frequency scale is
\begin{equation}
  \delta \omega_{\textrm{x}} \sim \frac{\sqrt{K}\gamma_c}{|\alpha(\omega_0)|}
  = \left[\frac{\hbar \omega_0 \Delta_{\IN}}{P_{\IN}}\right]^{1/2}\,
  \sqrt{K} \, \gamma_c.
\end{equation}
The measured values of the signal and noise must, however, be obtained
by averaging (\ref{sra}) over the bandwidth $\Delta_{\IN}$ at each
frequency, as indicated by the dashed curves in Fig.~\ref{fig_zerot}.
These averaged values do not vanish at $\omega = \omega_0$.  Up to a
factor of order unity depending on the averaging procedure, their
residual values are
\begin{equation}
  \overline{\braket{\N}}_{\omega_0} =
  \overline{\braket{\delta\N^2}}_{\omega_0} \approx
  \frac{P_{\IN}(\omega_0)\,\Delta_{\IN}}{12\hbar\omega_0 K \gamma_c^2}
  \sim \left[\frac{\Delta_{\IN}}{\delta \omega_{\textrm{x}}}\right]^2.
  \label{bandwidth average}
\end{equation}
In particular, we can regard
$[\overline{\braket{\delta\N^2}}_{\omega_0}]^{1/2}$ as the effective
shot noise level at the absorption resonance.  It dominates over the
bandwidth-averaged signal if $\Delta_\IN \ll
\delta\omega_{\textrm{x}}$.


The crossover frequency scale $\delta \omega_{\textrm{x}}$ is related
to the sensitivity of the output to the loss detuning parameter
$\delta\gamma$.  From Eq.~(\ref{T=0 CPA}), the change in the output
signal at $\omega = \omega_0$ resulting from $\delta\gamma \ne 0$ is
\begin{equation}
  \braket{\Delta\N} = \left[\frac{\delta\gamma}{\delta\omega_{\textrm{x}}}\right]^2.
  \label{change in output}
\end{equation}
(To lowest order, this quantity is unaffected by bandwidth-averaging.)
The sensitivity of the CPA as a low-background detector is given by
the minimum $\delta \gamma$ for which $\braket{\Delta\N}$ is distinguishable
from the effective noise level.  Comparing (\ref{change in output}) to
(\ref{bandwidth average}), we obtain the result
\begin{equation}
  |\delta\gamma| \gtrsim \sqrt{\delta\omega_{\textrm{x}} \Delta_{\IN}}.
\end{equation}

\begin{figure}
\centering
\includegraphics[width=0.38\textwidth]{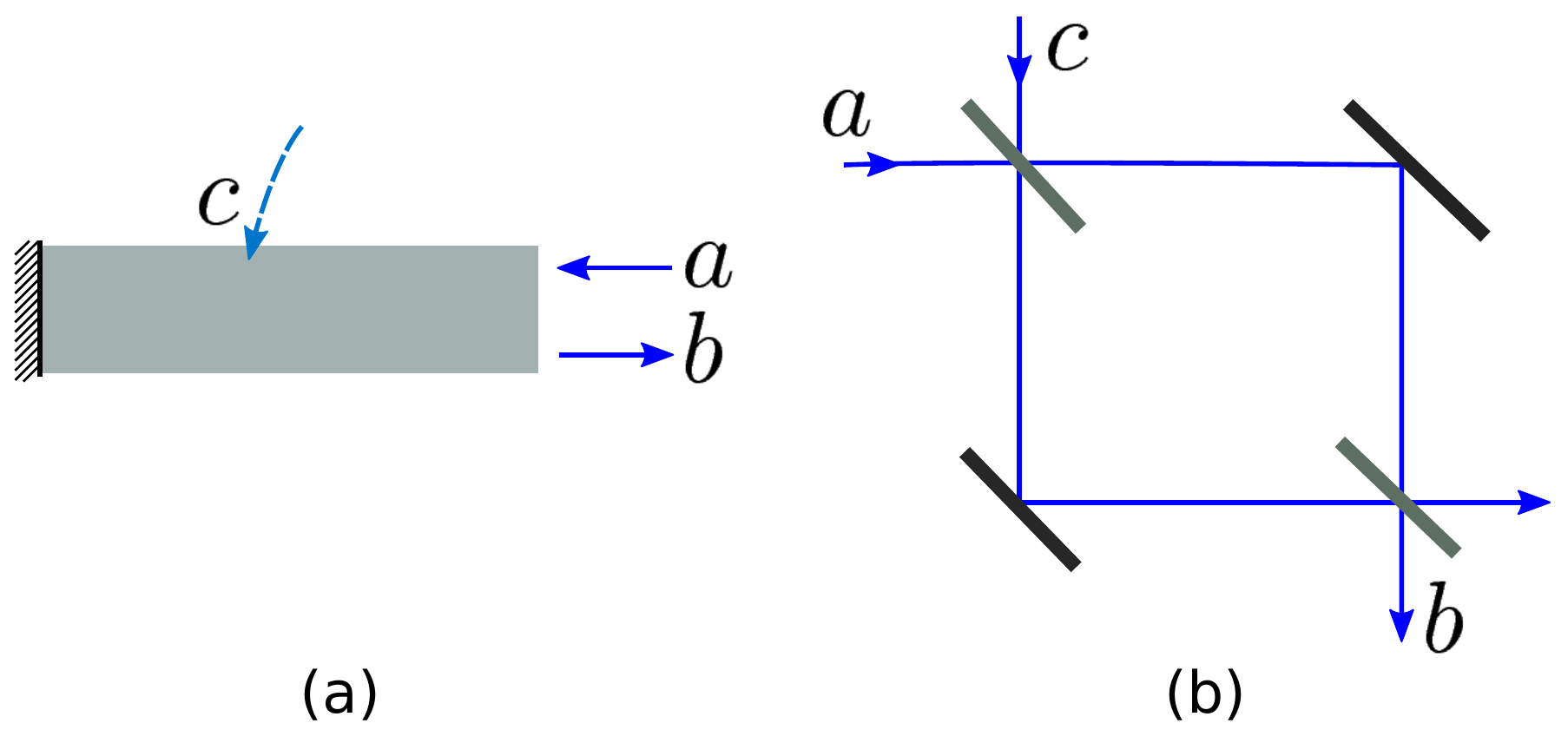}
\caption{(color online) (a) Schematic of a one-channel absorbing
  cavity, where the output amplitude $b$ is a superposition of the
  input amplitude $a$ and the reservoir operator $c$.  (b) The
  corresponding Mach-Zender interferometer, with $a$ and $c$ entering
  in the two input ports and $b$ being one of the outputs.}
\label{fig_mz}
\end{figure}

Since a CPA can serve as an absorbing interferometer/detector, it is
useful to compare it to a lossless interferometer such as a
Mach-Zender interferometer (MZ).  As shown in Fig.~\ref{fig_mz}, an analogy can be made
between a CCR (single-channel CPA) and a MZ.  The input photon operator
$a$ and the reservoir operator $c$, from the one-channel version of
Eq.~(\ref{input output}), map onto two input photon operators for the
MZ; meanwhile the output photon operator $b$ maps onto one of the MZ
outputs.  The fluctuation-dissipation relation,
Eq.~(\ref{fluct-diss}), is equivalent to the relation for transmission into the {\it observed port} of the MZ, $S_b$, (which follows from the 
unitary scattering matrix for both ports):
\begin{equation}
  S_b = \frac{1}{2}\left(e^{i\theta} - 1\right), \quad |U^2| = 1 -
  |S_b|^2,
\end{equation}
where $\theta$ is the phase difference between the two arms of the
interferometer.  Perfect absorption in the cavity corresponds to
$\theta = 0$, so that the $a$ input is directed entirely into the
second, unmonitored output port, leaving the $b$ port empty.  Quantum
fluctuations in the empty $c$ input contribute to the shot noise.
Note that in the CCR, the single input beam interferes with itself and
it does not function as an optical interferometer, i.e.~it does not
measure the relative phase of two input beams.  The multi-channel CPA
does, however, act as an interferometer.  In the simplest case of two
input channels and a cavity with parity symmetry, the two $S$-matrix
eigenmodes have even and odd symmetry respectively, and only one of
these are perfectly absorbed at a given absorption resonance.  The
interferometry is performed by making small phase changes between the
two input beams, moving the system slightly away from the
eigenchannel.  For $N > 2$ and/or absent parity symmetry, both the
magnitude and phase of each input amplitude must be tuned in order to
reach perfect absorption; this is analogous to a multi-input MZ-like
interferometer whose input amplitudes and phases may be tuned to send all the output into a
single port, playing the role of the absorbing channel.

The main difference between the CPA and the MZ lies in the frequency
dependence.  For the latter, $S_b$ varies sinusoidally in frequency with
free spectral range $\sim c / L$, where $L$ is the dimension of the
system.  Absorbing cavities, however, are described by Eq.~(\ref{s
  omega}), and for high-Q cavities the absorption resonances are much
narrower than $c / L$.  Since the signal-to-noise ratio depends
inversely on $\gamma_c, 1/L$ in the two cases, for a given value of
$\omega-\omega_0$, the CPA would have a better signal to noise ratio.
This resonant enhancement of the sensitivity of a CCR has been pointed
out in a different context in Ref.~\cite{Chow}. A similar effect could be achieved in
a MZ by adding resonant cavities along each arm to effectively increase the 
optical path length.

Finally, we make some comments about the $T > 0$ case.  In
Eq.~(\ref{total N output}), we see that $\braket{\N}$ is written as
the sum of the classical scattered flux and the gray-body thermal
emission, $ f(T)\, N [1-\bar\sigma]$, $\bar \sigma$ being the mean
scattering strength per channel \cite{beenakker0}.  For a CCR
(one-channel CPA), $\bar \sigma = 0$ at the operating frequency of the
absorption resonance, so the thermal emission has the black-body
value.  In the MZ analogy, Fig.~\ref{fig_mz}(b), this is equivalent to
connecting the $c$ input port to a black-body source.  Since the MZ is
tuned so that the $a$ input is completely directed into the
unmonitored output port, the black-body emission into $c$ is directed
into the monitored output port $b$.

For a multi-channel CPA, the thermal emissivity is less than the
black-body value, even at the operating frequency of the absorption
resonance.  This is because only one of the $N$ scattering strengths
vanishes; the other $N - 1$ scattering strengths are nonzero, so that
$\bar \sigma > 0$.  From Eq.~(\ref{variance vs mean}), the same is
true of the thermal contribution to the output noise $\braket{\N^2}$.
Furthermore, for $N \gg 1$ a ``hidden black'' scenario is possible, in
which the CPA perfectly absorbs the input field, $\alpha$, but emits a
negligible amount of thermal radiation \cite{hidden black}.  More
specifically, for a weakly absorbing system it is possible that the
mean albedo (reflectivity) can be large, $\bar \sigma \rightarrow 1$,
indicating an almost white body, and implying that the thermal
contributions to $\braket{\N}$ and $\braket{\N^2}$ vanish, while
nonetheless the system is perfectly absorbing (down to the quantum
noise floor) if the correct input field is supplied.  Elsewhere, two
of the authors have shown that this effect can be generalized beyond
the case of a perfect CPA at resonance, to a disordered scattering
medium with weak absorption over a large range of frequency and
absorptivity \cite{hidden black}.

The presence of thermal noise also affects the sensitivity of the CPA
as a detector.  From Eq.~(\ref{output noise}), the noise level at
$\omega = \omega_0$ in the thermal noise dominated limit is
\begin{equation}
  \braket{\delta\N^2}_{\omega_0}^{\textrm{thermal}} \equiv g(T)
  \approx f(T) \; \textrm{Tr}\left(\mathbf{1} - SS^\dagger\right).
\end{equation}
Comparing this to the signal (\ref{change in output}), we obtain the
sensitivity limit
\begin{equation}
  |\delta\gamma| \gtrsim g(T)^{1/4} \, \delta\omega_{\textrm{x}}.
\end{equation}
The crossover between the bandwidth-dominated and thermal-dominated
noise regimes occurs at
\begin{equation}
  g(T) \sim \left[\frac{\Delta_{\IN}}{\delta\omega_{\textrm{x}}}\right]^2.
\end{equation}

This research was supported by NSF ECCS grant 1068642, and by NRF
(Singapore) grant NRFF2012-02.  The authors would like to thank
M.~Devoret for helpful discussions.


\begin{thebibliography}{99}
\bibitem{note} The critically-coupled resonator and other CPA
  precursors correspond to the special case in which there is only a
  single input/output channel and the $S$-matrix degenerates to a
  number.
\end{thebibliography}
\end{document}